\documentclass[epjCONF,columns,a4paper]{svjour} 
\usepackage{graphics}
\usepackage{graphicx}  
\usepackage[varg]{txfonts} 
\usepackage[latin1]{inputenc}
\session-title{Hadron Collider Physics symposium (HCP-2011), Paris, France, November 14-18 2011}

\usepackage{xspace}
\usepackage{upgreek}
\def\CP                {\ensuremath{C\!P}\xspace}
\def\PD      {\ensuremath{D}\xspace}                 
\def\PK      {\ensuremath{K}\xspace}                 
\def\Ppi         {\ensuremath{\pi}\xspace}                 
\def\AP {\ensuremath{A_{\mathrm{P}}}\xspace}

\def\AD {\ensuremath{A_{\mathrm{D}}}\xspace}
\def\pis {\ensuremath{\pi_{\mathrm{s}}}\xspace}
\def\aindCP {\ensuremath{a^{\mathrm{ind}}_{\CP}}\xspace}
\def\adirCP {\ensuremath{a^{\mathrm{dir}}_{\CP}}\xspace}
\def\ARAW {\ensuremath{A_{\mathrm{raw}}}\xspace}
\def\kaon  {\ensuremath{\PK}\xspace}
\def\Kp    {\ensuremath{\kaon^+}\xspace}
\def\Km    {\ensuremath{\kaon^-}\xspace}
\def\Dbar    {\kern 0.2em\overline{\kern -0.2em \PD}{}\xspace}
\def\D       {\ensuremath{\PD}\xspace}
\def\Dz      {\ensuremath{\D^0}\xspace}
\def\Dzb     {\ensuremath{\Dbar^0}\xspace}
\def\Dstarp  {\ensuremath{\D^{*+}}\xspace}
\def\Dstarm  {\ensuremath{\D^{*-}}\xspace}

\def\pion  {\ensuremath{\Ppi}\xspace}
\def\pip   {\ensuremath{\pion^+}\xspace}
\def\pim   {\ensuremath{\pion^-}\xspace}
\def\pt         {\mbox{$p_{\rm T}$}\xspace}
\def\mum  {\ensuremath{\,\upmu\rm m}\xspace}

\begin{document}
\title{A search for time-integrated \CP violation in $\boldmath D^0 \to h^-h^+$ decays}
\author{M. J. Charles\inst{1}\fnmsep\thanks{\email{m.charles1@physics.ox.ac.uk}}, on behalf of the LHCb collaboration.}
\institute{Department of Physics, University of Oxford, Oxford, United Kingdom}
\abstract{
  The preliminary results of a search for time-integrated \CP violation in $\Dz \to h^-h^+$ ($h=K$, $\pi$) decays 
  performed with 0.6~fb$^{-1}$ of data collected by LHCb in 2011 are presented.
  The flavour of the charm meson is determined by the charge of the slow pion in the $\Dstarp \to \Dz \pip$ and $\Dstarm \to \Dzb \pim$ decay chains.
  The difference in \CP asymmetry between
  $\Dz \to K^- K^+$ and $\Dz \to \pi^- \pi^+$, $\Delta A_{\CP} \equiv  A_{\CP}(K^-K^+) \, - \, A_{\CP}(\pi^-\pi^+)$,  is measured to be
  $\Delta A_{\CP} = \left[ -0.82 \pm 0.21 (\mathrm{stat.}) \pm 0.11 (\mathrm{syst.}) \right]\%$.
  This differs from the hypothesis of \CP conservation by $3.5\sigma$. 
} 
\maketitle

%

\section{Introduction}

The time-integrated \CP asymmetry,  $A_{\CP}(f)$,  for the final states $f= K^- K^+$
and $f= \pi^- \pi^+$ has two contributions: 
  an indirect component (to a good approximation universal for \CP eigenstates in the Standard Model)
  and a direct component (in general final state dependent). In the
  limit of U-spin symmetry, the direct component is equal and opposite in sign 
for $K^-K^+$ and $\pi^-\pi^+$~\cite{bib:grossman_kagan_nir}.
In the Standard Model, \CP violation is expected to be small~\cite{bib:cicerone,bib:lenz,bib:grossman_kagan_nir}.
However, in the presence of physics beyond the Standard Model the rate of \CP violation
could be enhanced~\cite{bib:grossman_kagan_nir,bib:littlest_higgs}.
No prior evidence of \CP violation in the charm sector has been found.

The most precise measurements to date of the time-integrated \CP
asymmetries in $\Dz \to K^- K^+$ and $\Dz \to \pi^- \pi^+$ were made
by the CDF, BABAR, and BELLE
collaborations~\cite{bib:cdf_paper,bib:babar_paper2008,bib:belle_paper2008}.
LHCb has previously reported preliminary results based on 37~$\rm pb^{-1}$ of data collected in 2010~\cite{bib:lhcb2010}.
In the limit that the efficiency for selected events is independent of the decay time, the difference between
the two time-integrated asymmetries,
$\Delta A_{\CP} \equiv A_{\CP}(K^-K^+) \, - \, A_{\CP}(\pi^-\pi^+)$,
is equal to the difference in the direct \CP asymmetry. However, if the
dependence of the efficiency on the decay time is different for the $\Km\Kp$ and $\pim\pip$ final
states, a contribution from indirect \CP violation remains.
The Heavy Flavor Averaging Group (HFAG) has combined time-integrated
and time-dependent measurements of \CP asymmetries taking account of
the different decay time acceptance to obtain world-average
values for the indirect \CP asymmetry of $a_{\CP}^{\mathrm{ind}} = (-0.03 \pm 0.23)\%$ and
the difference in direct \CP asymmetry between the final states of $\Delta a_{\CP}^{\mathrm{dir}} = (-0.42 \pm 0.27)\%$~\cite{bib:hfag}.

In these proceedings, LHCb results~\cite{bib:this_result_conf,bib:this_result_eprint}
for the measurement
of the difference in integrated \CP asymmetries between $D^0 \to K^-K^+$
and $D^0 \to \pi^-\pi^+$, performed with approximately 0.6~$\rm fb^{-1}$
of data collected in 2011, are presented.
The initial state (\Dz or \Dzb) is tagged by
requiring a $\Dstarp \to \Dz \pip$ decay.  
The use of charge-conjugate modes is implied throughout,
except in the definition of asymmetries.

\section{Formalism}
\label{sec:formalism}

The raw asymmetry for tagged \Dz decays to a final state $f$
is given by $\ARAW(f)$, defined as:

\begin{equation}
\ARAW(f) \equiv \frac{N(D^{*+} \to D^0( f)\pi^+) \, - \, N(D^{*-} \to \Dzb (\bar{f})\pi^-)}
                                            {N(D^{*+} \to D^0( f)\pi^+) \, + \, N(D^{*-} \to \Dzb (\bar{f})\pi^-)},
\label{def:astarrawdef}
\end{equation}
where $N(X)$ refers to the number of reconstructed events of decay $X$
after background subtraction. 

The raw asymmetries may be written as a sum of various components,
coming from both physics and detector effects:
\begin{eqnarray}
\ARAW(f) &=& A_{\CP}(f) \, + \, \AD(f) \, + \, \AD(\pis) \, + \, \AP(D^{*+}).
\label{def:arawstarcomponents}
\end{eqnarray}
Here, $A_{\CP}(f)$ is the intrinsic physics \CP asymmetry,
$\AD(f)$ is the asymmetry for selecting the $D^0$ decay
into the final state $f$,  
$\AD(\pis)$ is the asymmetry for selecting the `slow pion'
from the $D^{*+}$ decay chain, and 
$\AP(D^{*+})$ is the production asymmetry
for prompt $D^{*+}$ mesons.
The asymmetries $A_{\CP}$, $\AD$ and $\AP$ are defined in
the same fashion as \ARAW.

For a two-body decay of a spin-0 particle to a self-conjugate
final state there can be no $D^0$ detection asymmetry,
i.e. $\AD(K^-K^+) = \AD(\pi^-\pi^+) = 0.$  Moreover, to first order $\AD(\pis)$ and $\AP(D^{*+})$ cancel out in the difference
\begin{displaymath} 
\ARAW(K^-K^+) \, - \, \ARAW(\pi^-\pi^+) ,
\end{displaymath}
leaving a quantity, defined as $\Delta A_{\CP}$, equal to the difference in physics asymmetries:
\begin{eqnarray}
\Delta A_{\CP} & \equiv & A_{\CP}(K^-K^+) \, - \, A_{\CP}(\pi^-\pi^+), \label{eq:deltaacpdef} \\
& = & \ARAW (K^-K^+) \, - \, \ARAW (\pi^-\pi^+).\label{eq:adefequals}
\end{eqnarray}
To minimize second order effects, related to the slightly different kinematic properties of the two decay modes, the analysis is done in bins
of the relevant kinematic variables, as shown later in Secs.~\ref{sec:massfits} and~\ref{sec:results}.
The physics asymmetry of each final state may be written at first order as~\cite{bib:bigi_d2hh}:
\begin{eqnarray}
A_{\CP}(f) &  \approx & \adirCP(f) \, + \, \frac {\langle t \rangle}{\tau} \aindCP, \label{eq:acpphysics}
\end{eqnarray}
where
$\adirCP(f)$ is the asymmetry coming from direct \CP violation for the decay,
$\langle t \rangle$ is the average decay time in the sample used, 
$\tau$ the true $D^0$ lifetime, and 
$\aindCP$ is the asymmetry associated with \CP violation in the mixing.
To a good approximation this latter quantity is
universal~\cite{bib:grossman_kagan_nir}, and so
\begin{eqnarray}
\Delta A_{\CP} &  = & \left[ \adirCP(K^-K^+) \,-\, \adirCP(\pi^-\pi^+) \right] \, + \, \frac {\Delta \langle t \rangle}{\tau} \aindCP, \label{eq:acpfinal}
\end{eqnarray}
where
$\Delta \langle t \rangle$ is the difference in average decay time
of the $D^0$ mesons in the $K^-K^+$ and $\pi^-\pi^+$ samples.
In the limit that  $\Delta \langle t \rangle$ vanishes,
$\Delta A_{\CP}$ probes the difference in direct \CP violation
between the two decays.   

\section{Dataset and selection}
\label{sec:dataset}

A description of the LHCb detector may be found in
Ref.~\cite{LHCb}.  The field direction in the LHCb dipole is such that charged 
particles are deflected in the horizontal plane.  The current direction in the dipole was changed
several times during data taking; about 60\% of the data was taken
with one polarity and 40\% with the other.

Selections are applied to provide samples of $D^{*+}\to D^0 \pi^+$ 
candidate decays, with $D^0 \to K^-K^+$ and $\pi^-\pi^+$.
A loose $D^0$ selection including a mass window of full width 100~MeV$/c^2$
was already applied during data taking, in the final stage of the
high level trigger (HLT).   
In the offline analysis only candidates that were accepted by this
trigger algorithm are considered.
Both the offline and HLT selections impose a variety of requirements on
kinematics and decay time to isolate the decays of interest, 
including requirements
  on the track fit quality,
  on the \Dz and \Dstarp vertex fit quality,
  on the transverse momentum of the \Dz ($\pt > 2$~GeV$/c$),
  on the decay time $t$ of the \Dz ($ct > 100 \, \mum$),
  on the helicity angle of the \Dz decay,
  that the \Dz points back to a primary vertex,
  and that the \Dz daughter tracks do not.
Fiducial requirements are imposed to ensure
  that the tagging soft pion lies within the central region of the detector acceptance.
In addition, the offline analysis exploits the capabilities of the RICH
system to distinguish between pions and kaons when reconstructing the \Dz.

\begin{figure}
  \begin{center}
    \includegraphics[width=\columnwidth]{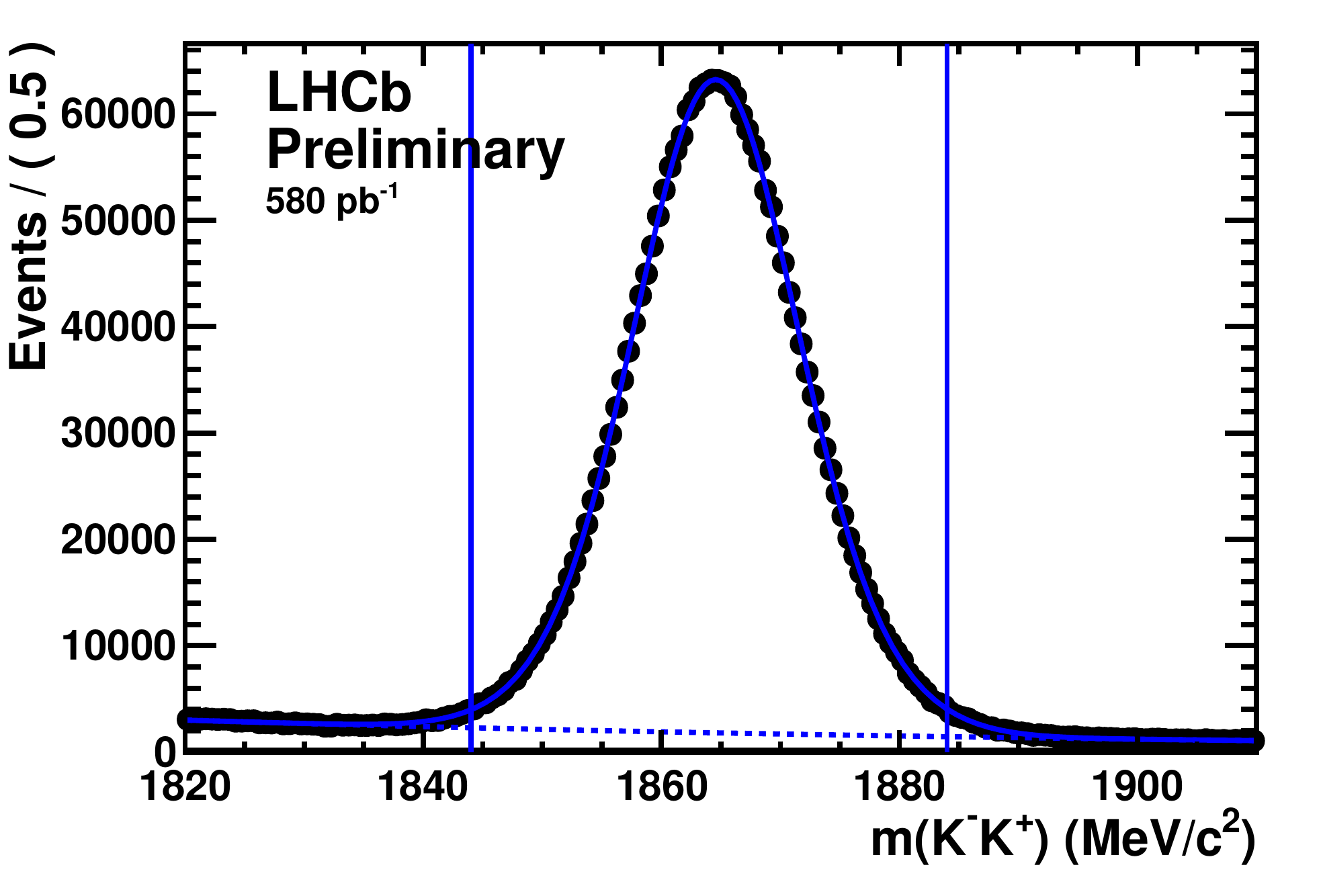}
    \includegraphics[width=\columnwidth]{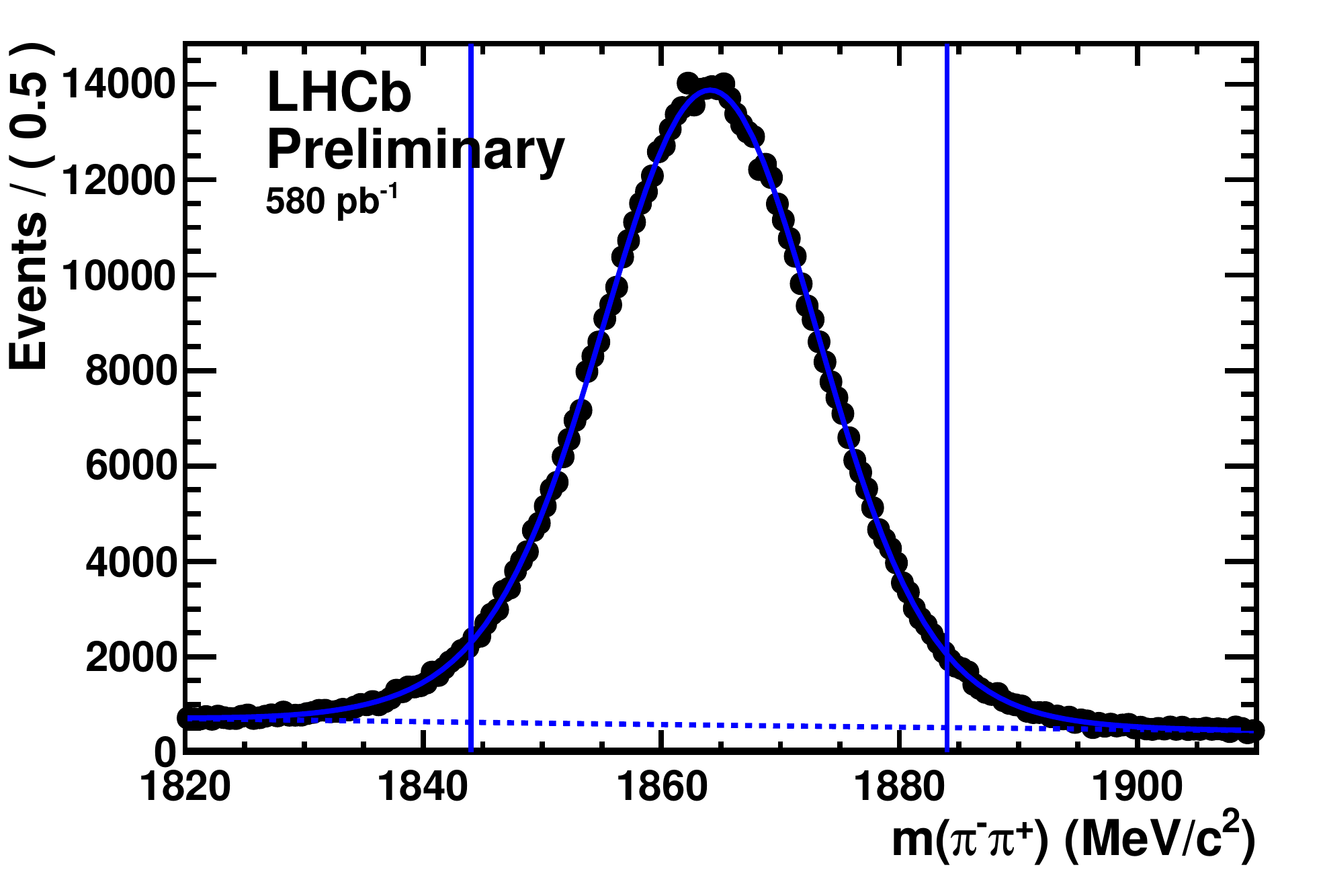}
   \end{center}
  \caption{
    Fits to the $m(\Km\Kp)$ and $m(\pim\pip)$ spectra of \Dstarp candidates passing
    the selection and satisfying $0<\delta m<15$~MeV$/c^2$. The vertical blue lines indicate
    the signal window of 1844--1884~MeV$/c^2$.
  }
  \label{fig:massTagged}
\end{figure}

\begin{figure}
  \begin{center}
    \includegraphics[width=\columnwidth]{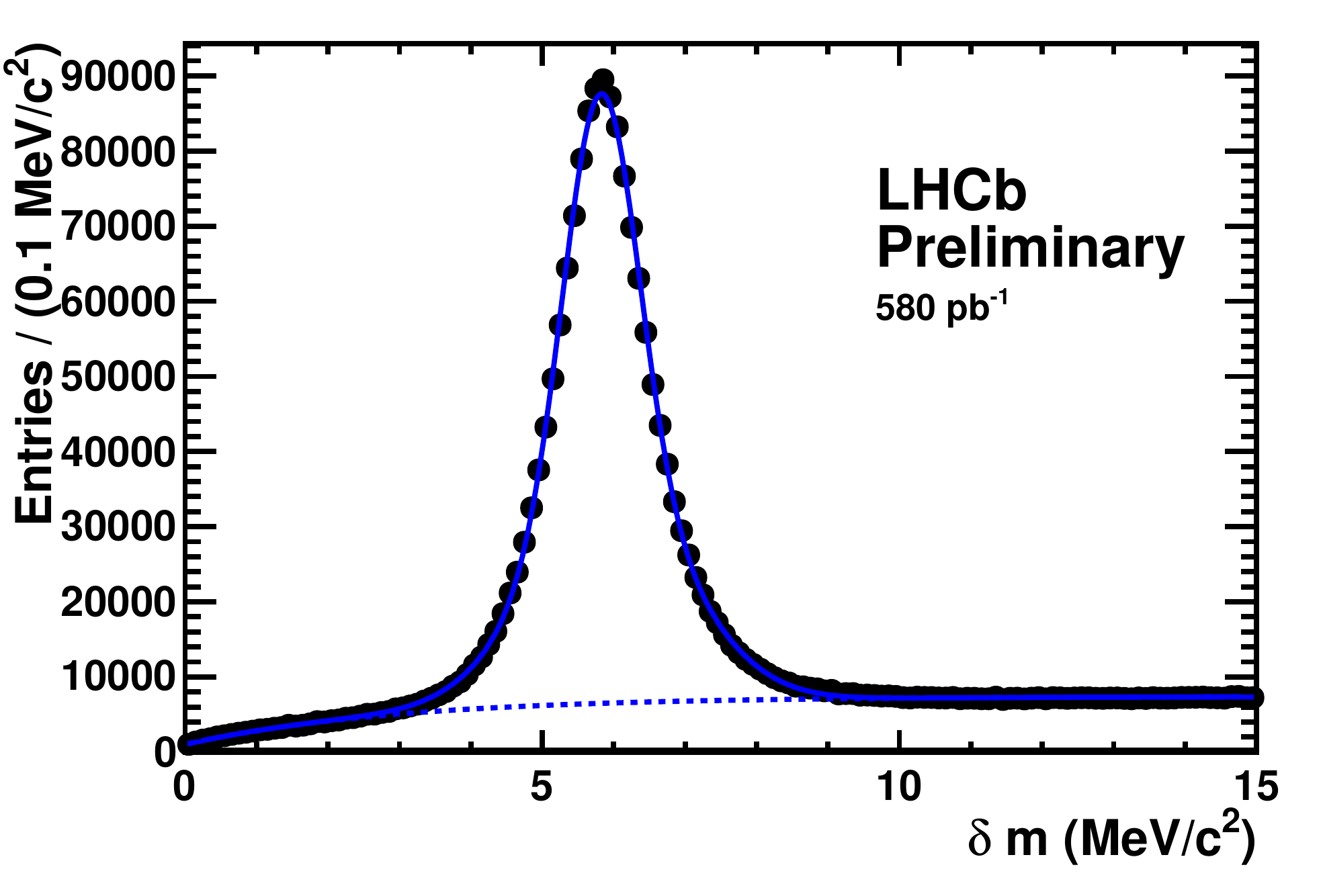}
    \includegraphics[width=\columnwidth]{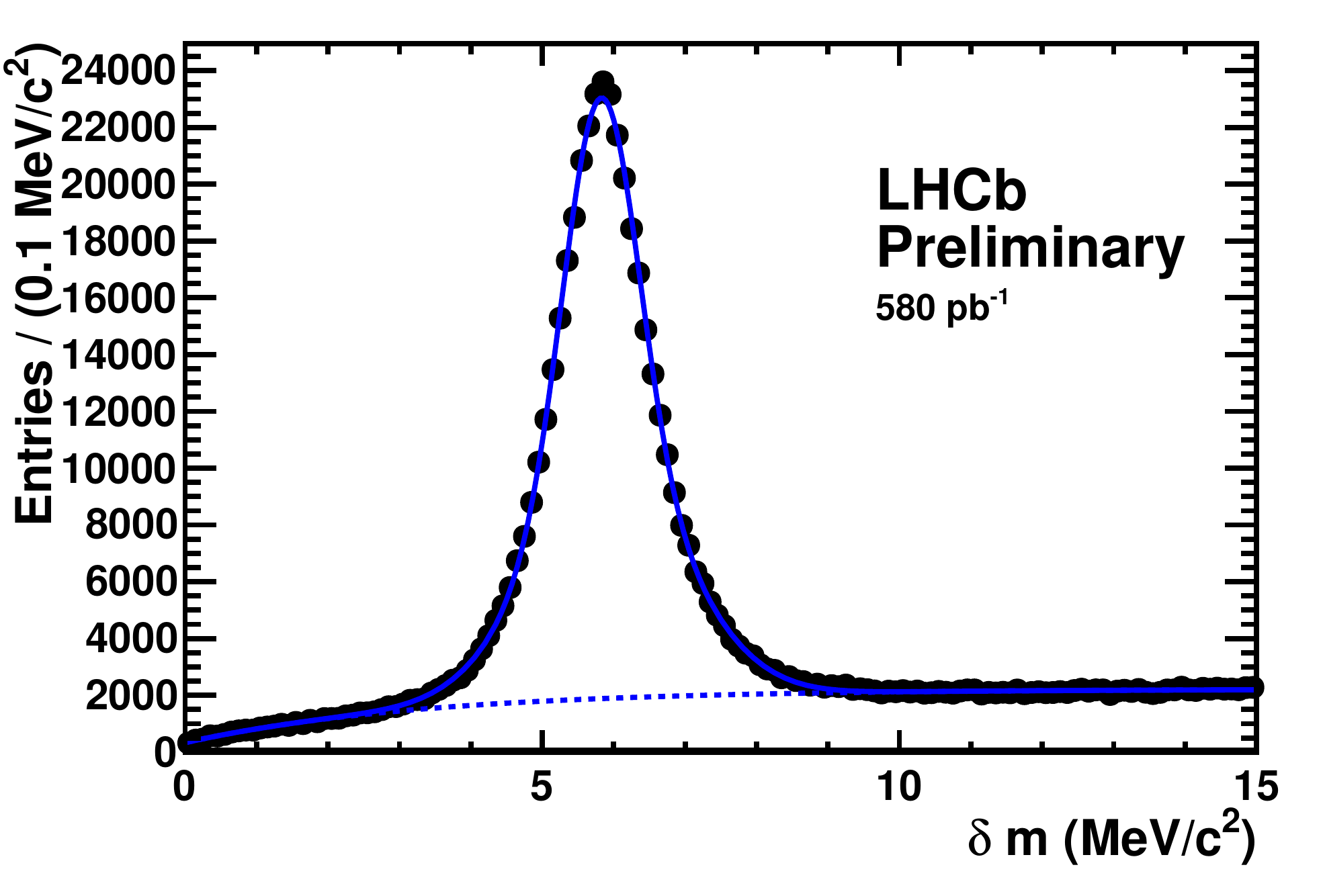}
   \end{center}
  \caption{
    Fits to the mass difference spectra, where the $D^0$ is reconstructed
    in the final states $K^- K^+$ (top) and $\pi^- \pi^+$ (bottom), with a $\Dz$ mass lying in the window of 1844--1884~MeV$/c^2$.
  }
  \label{fig:deltaMassTagged}
\end{figure}

Defining the mass difference as 
$\delta m \equiv m(h^+ h^- \pi^+) - m(h^+ h^-) - m(\pi^+)$,
the mass and mass difference spectra of selected candidates
are shown in 
Figs.~\ref{fig:massTagged} and~\ref{fig:deltaMassTagged}, respectively.
The difference in width between the $K^-K^+$ and $\pi^-\pi^+$ mass spectra arises from 
the different opening angles of the two decays.
The \Dstarp signal yields are approximately
$1.44 \times 10^6$ in the tagged $K^-K^+$ sample,
and $0.38 \times 10^6$ in the tagged $\pi^-\pi^+$ sample.
The  fractional difference in average decay time of $D^0$ candidates passing the selection
between the $K^-K^+$ and $\pi^-\pi^+$ samples is
$\Delta \langle t \rangle / \tau = (9.8 \pm 0.9)\%$.

\section{Fit procedure}
\label{sec:massfits}

Fits are performed on the samples in order to determine
$\ARAW(K^-K^+)$ and $\ARAW(\pi^-\pi^+)$.
The analysis is performed in 54 kinematic bins, divided
by \pt and pseudorapidity ($\eta$) of the $D^{*+}$ candidates,
momentum of the tagging soft pion, and whether the initial trajectory of the slow pion is 
towards the left or right half of the detector.
A binning is imposed for the reason that the production and detection 
asymmetries can in general vary with \pt and $\eta$, and so can the 
detection efficiency of the two different $D^0$ decays, in particular 
through the effects of the particle identification requirements.  

The events are further partitioned in two ways.
First, the data are divided between the two dipole magnet settings.
Second, the first 350~pb$^{-1}$ of data are processed separately from
the remainder, with the division aligned with a
break in data taking due to a LHC technical stop.
In total, therefore, 216 independent measurements are considered for each decay mode.

\begin{figure}
  \begin{center}
    \includegraphics[width=\columnwidth]{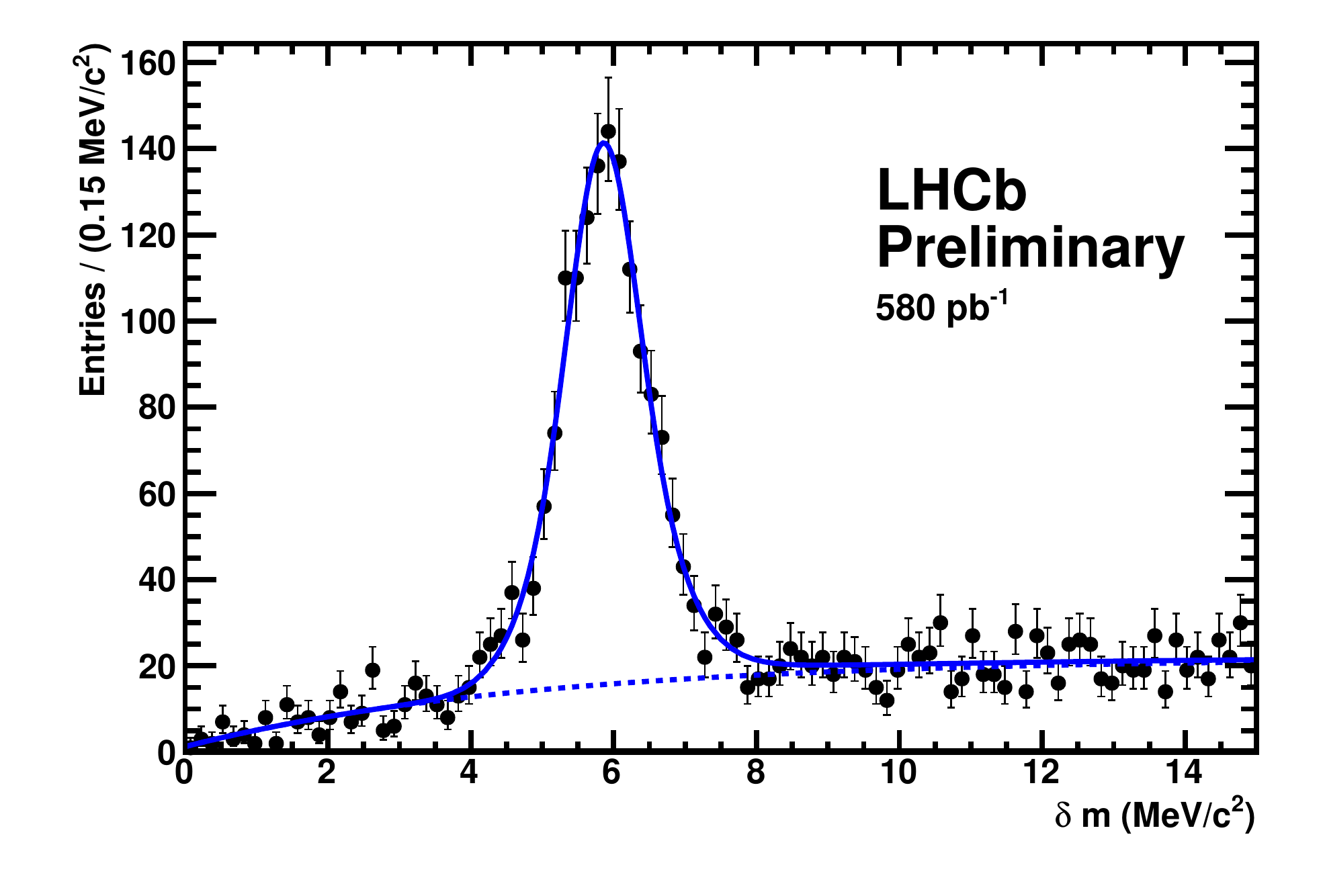}
   \end{center}
  \caption{
    Example fit used in the $\Delta A_{CP}$ analysis. The first kinematic
    bin of the first run period with magnet up polarity is shown for the
    $D^0 \to K^-K^+$ final state.
  }
  \label{fig:exampleFit}
\end{figure}

One-dimensional fits to the mass difference spectra are performed.
The fits include separate components for the signal and background lineshapes.
The signal is described as the sum of two Gaussian functions with a common
mean, convolved with an asymmetric function.
The background is described by an empirical function of the form
\begin{displaymath}
\left[ 1 - e^{-(\delta m - \delta m_0)/c} \right] ,
\end{displaymath}
where $\delta m_0$ and $c$ are parameters describing the threshold and shape of the
function, respectively. In each case an unbinned maximum likelihood
fit is used. The \Dstarp and \Dstarm samples are fitted simultaneously
and share several shape parameters, though a charge-dependent offset in the central
value and overall scale factor in the mass resolution are allowed. The raw asymmetry in the
yields of the signal component is extracted directly from this simultaneous fit.
An example fit from one measurement bin  is shown in Fig.~\ref{fig:exampleFit}.

The one-dimensional fits used for the tagged data do not distinguish between
correctly reconstructed signal and backgrounds that peak in
the mass difference. Such backgrounds can arise from \Dstarp decays in which the correct slow pion is found but the \Dz is partially mis-reconstructed. However, these backgrounds are suppressed by the use of tight
particle identification requirements and a narrow $D^0$ mass window. From studies of
the $D^0$ mass sidebands this contamination is found to be approximately 1\% of the signal yield.

\section{Results and systematic uncertainties}
\label{sec:results}

A value of $\Delta A_{\CP}$ is determined in each measurement bin using the
results for $\ARAW(K^-K^+)$ and $\ARAW(\pi^-\pi^+)$.   The $\chi^2/ndf$ of these measurements has a value of 211/215.
A weighted average is performed to yield the result $\Delta A_{\CP} =  (-0.82 \pm 0.21 )\%$.

\begin{figure}
  \begin{center}
    \includegraphics[width=\columnwidth]{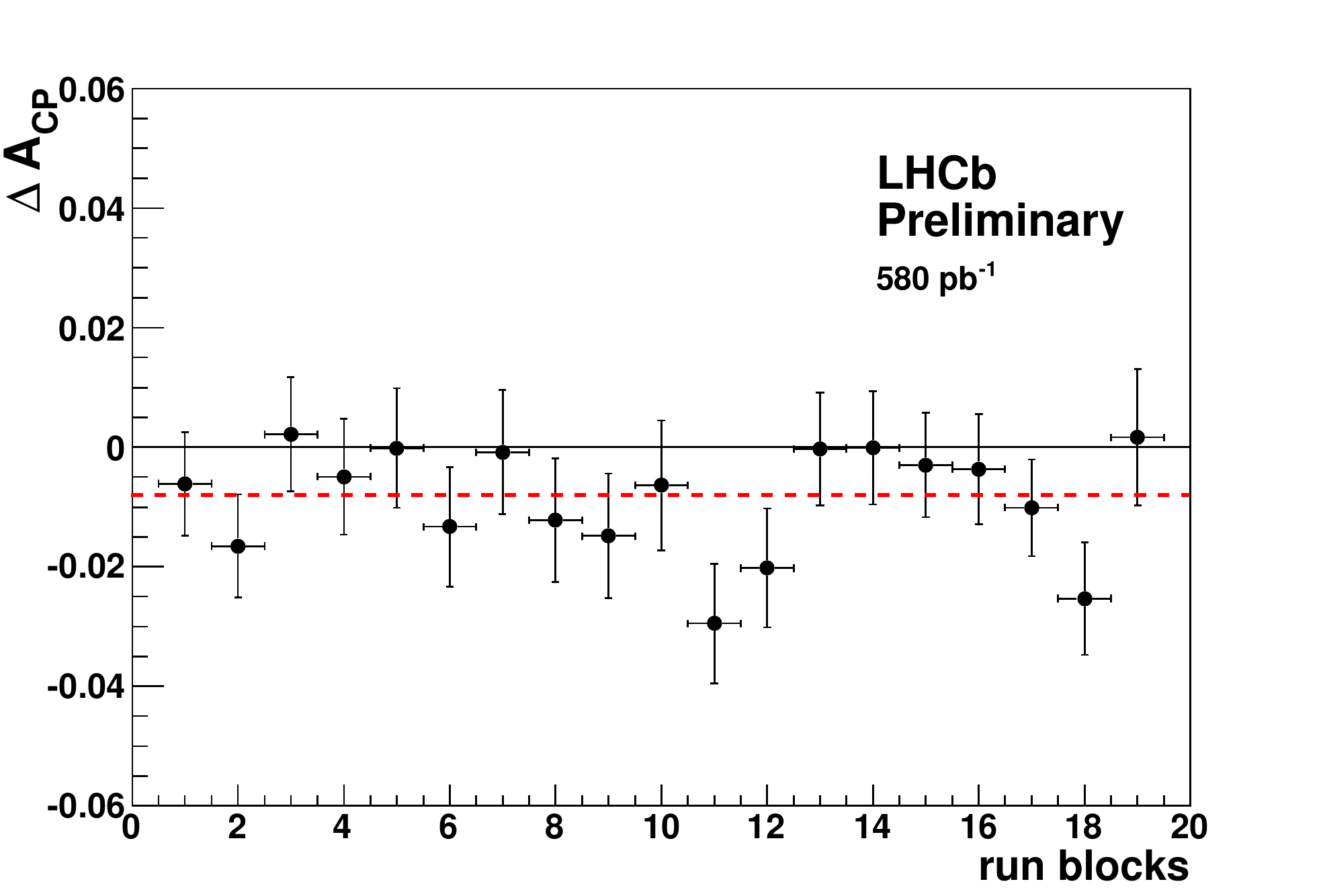}
   \end{center}
  \caption{
    Time-dependence of the measurement. The data are divided into
    19 disjoint, contiguous, time-ordered blocks and the value of
    $\Delta A_{CP}$ measured in each block. The red dashed line shows
    the result for the combined sample.
  }
  \label{fig:time}
\end{figure}

Numerous robustness checks are made, including
  monitoring the value of $\Delta A_{\CP}$ as a function of time (Fig.~\ref{fig:time}),
  re-performing the measurement with more restrictive RICH particle identification requirements, 
  and using a different $D^{*+}$ selection.
Potential biases due to the inclusive hardware trigger selection are investigated
with the subsample of data in which one of the signal final-state tracks is directly
responsible for the hardware trigger decision. In all cases good stability is observed. 

Systematic uncertainties are assigned 
  by loosening the fiducial requirement on the soft pion;
  by assessing the effect of potential effect peaking background in toy Monte Carlo studies;
  by repeating the analysis with the asymmetry extracted through sideband subtraction instead of a fit;
  with all candidates but one (chosen at random) removed in events with multiple candidates;
  and comparing with the result obtained with no kinematic binning.
In each case the full value of the change in result is taken as
the systematic uncertainty. These uncertainties are listed in
Table~\ref{tab:systematics:summary_acp_kk_pp}.  The sum in quadrature is $0.11\%$.

\begin{table}
  \caption{
    Summary of absolute systematic uncertainties for $\Delta A_{\CP}$.
  }
  \begin{center}
    \begin{tabular}{lc}
      \hline
      Effect & Uncertainty \\ 
      \hline
      Fiducial requirement & 0.01\% \\
      Peaking background asymmetry & 0.04\% \\
      Fit procedure & 0.08\% \\
      Multiple candidates & 0.06\% \\
      Kinematic binning & 0.02\% \\
      \hline
      Total & 0.11\% \\
      \hline
    \end{tabular}
  \end{center}
  \label{tab:systematics:summary_acp_kk_pp}
\end{table}

\section{Conclusions}

LHCb has measured the time-integrated difference in \CP asymmetry between $D^0 \to K^-K^+$ and $D^0 \to \pi^-\pi^+$ decays,
 $\Delta A_{\CP} =  \left( \adirCP(K^-K^+) \,-\, \adirCP(\pi^-\pi^+) \right) \, + \, 0.098 \, \aindCP$, to be
\begin{displaymath}
  \Delta A_{\CP} = \left[ -0.82 \pm 0.21 (\mathrm{stat.}) \pm 0.11 (\mathrm{syst.}) \right]\%
\end{displaymath}
with 0.6~fb$^{-1}$ of 2011 data,
where the first uncertainty is statistical and the second systematic.
Combining the statistical and systematic uncertainties in quadrature,
the significance of the measured deviation from zero is $3.5\sigma$.
The result is consistent with the current
HFAG world average~\cite{bib:hfag}.

Subsequent to this result being presented, several papers on the
theoretical interpretation have been written, examining the
possible size and uncertainty in the SM contribution and considering
possible new physics contributions. A partial list may be found in
Ref.~\cite{bib:reaction:Isidori,bib:reaction:Brod,bib:reaction:Wang,bib:reaction:Rozanov,bib:reaction:Pirtskhalava,bib:reaction:Hochberg,bib:reaction:Cheng,bib:reaction:Bhattacharya,bib:reaction:Chang}.
We thank our theory colleagues for their help in understanding this result.

\section*{Acknowledgements}

\noindent We express our gratitude to our colleagues in the CERN accelerator
departments for the excellent performance of the LHC. We thank the
technical and administrative staff at CERN and at the LHCb institutes,
and acknowledge support from the National Agencies: CAPES, CNPq,
FAPERJ and FINEP (Brazil); CERN; NSFC (China); CNRS/IN2P3 (France);
BMBF, DFG, HGF and MPG (Germany); SFI (Ireland); INFN (Italy); FOM and
NWO (The Netherlands); SCSR (Poland); ANCS (Romania); MinES of Russia and
Rosatom (Russia); MICINN, XuntaGal and GENCAT (Spain); SNSF and SER
(Switzerland); NAS Ukraine (Ukraine); STFC (United Kingdom); NSF
(USA). We also acknowledge the support received from the ERC under FP7
and the Region Auvergne.


\begin{thebibliography}{}
\bibitem{bib:grossman_kagan_nir}
  Y.~Grossman, A.~L.~Kagan and Y.~Nir,
  Phys.\ Rev.\  D {\bf 75} (2007) 036008.
\bibitem{bib:cicerone}
  S.~Bianco, F.~L.~Fabbri, D.~Benson and I.~Bigi,
  Riv.\ Nuovo Cim.\  {\bf 26N7} (2003) 1.
\bibitem{bib:lenz}
  M.~Bobrowski, A.~Lenz, J.~Riedl and J.~Rohrwild,
  JHEP {\bf 1003} (2010) 009.
\bibitem{bib:littlest_higgs}
  I.~I.~Bigi, M.~Blanke, A.~J.~Buras and S.~Recksiegel,
  JHEP {\bf 0907} (2009) 097.
\bibitem{bib:cdf_paper}
  CDF Collaboration,
  T.~Aaltonen {\it et al.},
  arXiv:1111.5023 [hep-ex] (submitted to Phys. Rev. D)
\bibitem{bib:babar_paper2008}
  BABAR Collaboration,
  B.~Aubert {\it et al.},
  Phys.\ Rev.\ Lett.\  {\bf 100 } (2008)  061803.
\bibitem{bib:belle_paper2008}
  Belle Collaboration,
  M.~Staric {\it et al.},
  Phys.\ Lett.\  B {\bf 670} (2008) 190.
\bibitem{bib:lhcb2010}
LHCb Collaboration, LHCb-CONF-2011-023.
\bibitem{bib:hfag}
  Heavy Flavor Averaging Group,
  D.~Asner {\it et al.},
  arXiv:1010.1589 [hep-ex];
  http://www.slac.stanford.edu/xorg/hfag/\\charm/EPS11/DCPV/direct\_indirect\_cpv.html.
\bibitem{bib:this_result_conf}
  LHCb Collaboration, LHCb-CONF-2011-061.
\bibitem{bib:this_result_eprint}
  LHCb Collabroation,
  R.~Aaij {\it et al.},
  arXiv:1112.0938 [hep-ex] (submitted to Phys. Rev. Lett.).
\bibitem{bib:bigi_d2hh}
  I.~I.~Bigi, A.~Paul and S.~Recksiegel,
  JHEP {\bf 1106} (2011) 089.
\bibitem{LHCb}
  LHCb Collaboration,
  A.~A.~Alves {\it et al.},
  JINST {\bf 3} (2008) S08005.
\bibitem{bib:reaction:Isidori}
  G.~Isidori, J.~F.~Kamenik, Z.~Ligeti and G.~Perez,
  arXiv:1111.4987 [hep-ph].
\bibitem{bib:reaction:Brod}
  J.~Brod, A.~L.~Kagan and J.~Zupan,
  arXiv:1111.5000 [hep-ph].
\bibitem{bib:reaction:Wang}
  K.~Wang and G.~Zhu,
  arXiv:1111.5196 [hep-ph].
\bibitem{bib:reaction:Rozanov}
  A.~N.~Rozanov and M.~I.~Vysotsky,
  arXiv:1111.6949 [hep-ph].
\bibitem{bib:reaction:Pirtskhalava}
  D.~Pirtskhalava and P.~Uttayarat,
  arXiv:1112.5451 [hep-ph].
\bibitem{bib:reaction:Hochberg}
  Y.~Hochberg and Y.~Nir,
  arXiv:1112.5268 [hep-ph].
\bibitem{bib:reaction:Cheng}
  H.-Y.~Cheng and C.-W.~Chiang,
  arXiv:1201.0785 [hep-ph].
\bibitem{bib:reaction:Bhattacharya}
  B.~Bhattacharya, M.~Gronau and J.~L.~Rosner,
  arXiv:1201.2351 [hep-ph].
\bibitem{bib:reaction:Chang}
  X.~Chang, M.-K.~Du, C.~Liu, J.-S.~Lu and S.~Yang,
  arXiv:1201.2565 [hep-ph].


\end{thebibliography}
\end{document}